\documentclass[fleqn,10pt]{wlscirep}
\usepackage[utf8]{inputenc}
\usepackage[T1]{fontenc}
\usepackage{graphicx}
\usepackage{subcaption}
\usepackage[utf8]{inputenc}
\usepackage[export]{adjustbox}
\usepackage{wrapfig}
\usepackage{tikz}
\usepackage{hyperref}
\usepackage{xcolor}
\usepackage[normalem]{ulem}
\title{Microwave response of a metallic superconductor subject to a high-voltage gate electrode}

\author[1,$\dagger$,*]{Giacomo Catto}
\author[1,2,$\dagger$,*]{Wei Liu}
\author[1]{Suman Kundu}
\author[1]{Valtteri Lahtinen}
\author[3]{Visa Vesterinen}
\author[1,3]{Mikko M\"ott\"onen}
\affil[1]{QCD Labs, QTF Centre of Excellence, Department of Applied Physics, Aalto University, P.O. Box 13500, FIN-00076
Aalto, Finland.}
\affil[2]{IQM, Keilaranta 19, 02150 Espoo, Finland}
\affil[3]{QTF Centre of Excellence, VTT Technical Research Centre of Finland Ltd,P.O. Box 1000, FI-02044 VTT, Finland}
\affil[$\dagger$]{These authors contributed equally.}
\affil[*]{Corresponding author email:giacomo.catto@aalto.fi; wei@meetiqm.com}

\begin{abstract}
Processes that lead to the critical-current suppression and change of impedance of a superconductor under the application of an external voltage is an active area of research, especially due to various possible technological applications. In particular, field-effect transistors and radiation detectors have been developed in the recent years, showing the potential for precision and sensitivity exceeding their normal-metal counterparts. In order to describe the phenomenon that leads to the critical-current suppression in metallic superconducting structures, a field-effect hypothesis has been formulated, stating that an electric field can penetrate the metallic superconductor and affect its characteristics. The existence of such an effect would imply the incompleteness of the underlying theory, and hence indicate an important gap in the general comprehension of superconductors. In addition to its theoretical value, a complete understanding of the phenomenon underneath the electric-field response of the superconductor is important in the light of the related technological applications. In this paper, we study the change of the characteristics of a superconductor implementing a coplanar-waveguide resonator as a tank circuit, by relating our measurements to the reactance and resistance of the material. Namely, we track the state of the superconductor at different voltages and resulting leakage currents of a nearby gate electrode which is not galvanically connected to the resonator. By comparing the effects of the leakage current and of a change in the temperature of the system, we conclude that the observed behaviour in the superconductor is mainly caused by the heat that is deposited by the leakage current, and bearing the experimental uncertainties, we are not able to observe the effect of the applied electric field in our sample. In addition, we present a relatively good quantitative agreement between the Mattis--Bardeen theory of a heated superconductor and the experimental observations. Importantly, we do not claim this work to nullify the results of previous works, but rather to provide inspiration for future more thorough experiments and analysis using the methods presented here.

\end{abstract}
\begin{document}

\flushbottom
\maketitle

\thispagestyle{empty}

\section*{Introduction}

Superconductors are an important piece of present and future technologies \cite{seidel2015applied}. In particular, superconducting radiation detectors\cite{yang2009suppressed,paiella2016development,wandui2020thermal,mazin2020superconducting,Kokkoniemi2020} and superconducting transistors \cite{ritter2021superconducting,Casparis2018} have been proposed to play an important role in diverse fields ranging from astronomy \cite{szypryt2017large,meeker2018darkness,alessandro2016lumped} to quantum computing \cite{Casparis2018,hernandez2020interplay,de2020niobium,zgirski2020heat,Larsen2015,deLange2015}. Moreover, such devices can be utilized in the study of the changes of the characteristics of a superconductor under an externally applied electric field: it is possible to detect the change of resistance and inductance caused by the field and control the critical current of the superconductor by applying an external voltage.

The mechanisms leading to the changes in the state of the superconductor have been under intensive investigation in the recent years. It is well understood that in semiconductor\cite{Clark1980, Takayanagi1985, Kleinsasser1989, Larsen2015, deLange2015, Casparis2018, massarotti2020high} and graphene\cite{Vora2012,Du2014, Kokkoniemi2020} devices the applied electric field penetrates into the material, leading to a change of the carrier density and hence the superconductor energy gap and critical current. This has enabled, e.g., the development of gate-tunable transmon qubits, referred to as gatemons,\cite{Casparis2018,Larsen2015,deLange2015} and a bolometer operating at the threshold for circuit quantum electrodynamics\cite{Kokkoniemi2020}.

In metallic devices however, the field effect is typically taken to be extinguished by the high density of the carriers and consequent efficient screening of the field inside the material. Nevertheless, major changes in the critical currents of metallic nanowires subject to different voltages on nearby gate electrodes have been reported\cite{paolucci2018ultra,de2018metallic,paolucci2019magnetotransport,alegria2021high,golokolenov2021origin}. In order to explain this behaviour of metallic superconductors, two different explanations have been formulated: whereas some studies \cite{ritter2021superconducting,golokolenov2021origin,alegria2021high} attribute it to heating arising from the injection of electric current from the insulated gate electrode to the superconductor, and hence a change of its characteristic properties, others \cite{de2018metallic,paolucci2019magnetotransport} propose the response of the superconductor to be a result of a field effect which is not predicted by the existing theories, calling in such a case for new physics. 

In this paper, we connect the characteristics of a metallic superconductor to the resonance frequency and internal quality factor of a superconducting coplanar-waveguide (CPW) resonator. This allows us to study the state of the superconductor under different perturbations and compare the responses. Thanks to the two distinct measurable quantities defining the state, the quality factor and the resonance frequency, we draw traces in this two-dimensional state space. Our key observation is that identical traces in the state space are obtained for a varying gate voltage that results in leakage current and a varying bath temperature. We thus suggest that, in our sample and experimental setup, any possible effect caused by the application of the external voltage is negligible in comparison to the behaviour that arises from the heating of the superconductor or equivalently from the excitation of quasiparticles. In addition, we obtain a relatively good agreement between the Mattis--Bardeen theory of a heated superconductor and our experimental observations.

\section*{Results}
\subsection*{Definition of the essential state of the superconductor}

The sample structure is shown in Fig.\ref{fig:sample_design} where a quarter-wavelength CPW resonator made of niobium is terminated to ground via an aluminum strip. Starting from the geometrical parameters of the design it is possible to estimate, for the lumped-element model, the geometric inductance $L_\mathrm{g}$, which consists of an external and internal contribution\cite{duzer1998principles}, and the corresponding capacitance $C$ of the fundamental mode of the CPW resonator \cite{goppl2008coplanar}, and hence the expected resonance frequency in the lumped-element model as
\begin{equation}\label{resonanceFrequency}
f=\frac{1}{2\pi\sqrt{L(L_\mathrm{g}, L_\mathrm{k}) C}}
\end{equation}
knowing that the total lumped-element-model inductance $L=L_\mathrm{g}+8 L_\mathrm{k}/\pi^2$ is supposed to change with the external perturbation due to the corresponding change of the aluminum kinetic inductance, $L_\mathrm{k}$. In the simplistic picture where the current density is assumed homogeneous across the cross-sectional area $A$ of the superconductor, we have\cite{Meservey1969}
\begin{equation}\label{kineticInductance}
L_\mathrm{k}=\frac{m}{e^2}\frac{V}{A^2}\frac{1}{n_\mathrm{s}} .
\end{equation}
Here, $m$ is the electron mass, $e$ is the elementary charge, $V$ is the volume of the superconductor, and $n_\mathrm{s}$ is twice the  density of Cooper pairs. The energy losses of the resonator are quantified by the total loaded quality factor\cite{PozarBook} 
\begin{equation}\label{qualityFactor}
Q=\left( \frac{1}{Q_\mathrm{i}}+\frac{1}{Q_\mathrm{ext}}\right)^{-1}
\end{equation}
where the external quality factor $Q_\mathrm{ext}$ corresponds to the losses due to the coupling of the resonator with its measurement setup and the internal quality factor $Q_\mathrm{i}$ takes into account the energy lost in other channels.
The latter can be expressed as a function of the lumped-element-model inductance, capacitance, and resistance of the resonator mode as
\begin{equation}\label{internalQ}
Q_\mathrm{i}=\frac{1}{R}\sqrt{\frac{L(L_\mathrm{g}, L_\mathrm{k})}{C}}
\end{equation}
relating it to the external perturbation through changes in these parameters. With equations~\eqref{resonanceFrequency} and \eqref{internalQ}, we thus have two characteristic properties of the resonator that are functions of the two parameters that characterise the state of the superconductor: the resistance and the inductance. Thus, in this work we identify each state of the superconductor by a distinct point in the $Q_\mathrm{i}-f$ plane, and observe its behavior under an external perturbation. Of course, the full microscopic description of the state of the superconductor requires more than just two parameters, but here we restrict our study to the framework of linear electric response. Note that in most of the previous studies, only the critical current, i.e., the inductance, of the superconductor has been studied. 

At finite temperature we expect the current in the superconductor to be carried by both, Cooper pairs and thermal quasiparticles. The density of the latter depends on the quasiparticle temperature $T$ according to \cite{de2011number}
\begin{equation}\label{n_q}
n_\textrm{q}(T) = 4N_0 \int_\Delta^\infty \frac{E\,\textrm{d}E}{\sqrt{E^2 -\Delta^2}\{1 +\exp{[E/(k_\textrm{B}T)]}\}} \simeq2N_0\sqrt{2\pi\Delta  k_\textrm{B} T}e^{-\Delta/(k_\textrm{B} T)} 
\end{equation}
where $N_0$ is the electron density of states at the Fermi level, $E$ is energy, and $\Delta=\Delta(T)$ is the energy gap parameter. The total density of charge carriers is given by the sum of the normal and superconductor contributions $n_s + n_q$. In the temperature range $T \ll \Delta(0)/k_\textrm{B}$ the thermal quasiparticle density is negligible; the temperature at which the thermal quasiparticles start to have a significant effect on the total density depends on the material, on the design of the sample, and on the experimental setup. There is a non-vanishing residual quasiparticle density even in the zero-temperature limit\cite{de2011number}, but such density is not expected to depend strongly on the perturbations we utilize. Thus, we are not expecting any significant change in the state of the resonator when considering low temperatures and no other perturbations.

\subsection*{Experimental setup}

\begin{figure}
\centering
\includegraphics[width=0.8\textwidth]{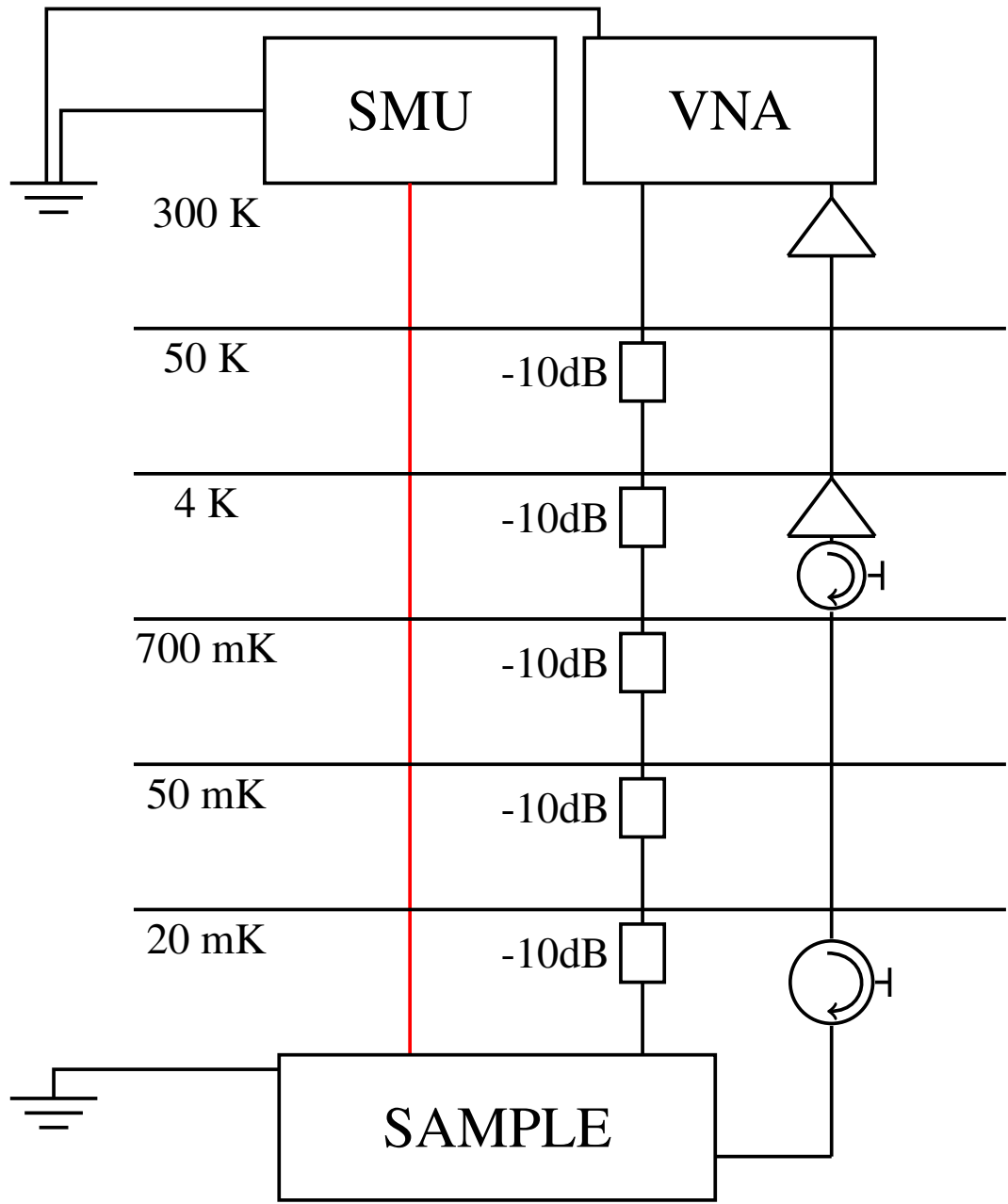}
\caption{Electrical schematic of the experimental setup. The red line represents a Thermocoax cable. The two grounds represent the same microwave ground, to which the direct current supply, the VNA, and the ground plane of the sample are connected.}
\label{fig:elec_scheme}
\end{figure}

Figure~\ref{fig:elec_scheme} shows the electrical schematic of the setup.
We track the changes of the inductance and resistance of the superconducting aluminium strip under the selected perturbations using the CPW resonator in a dilution refrigerator with a base temperature of 20~mK.
In the case of temperature sweeps, we use a built-in resistive heater to gradually increase the temperature of the mixing chamber plate where the sample is mounted from roughly $20$~mK up to $1.2$~K and carry out the measurement after giving the system enough time to attain thermal equilibrium. For gate voltage sweeps, we use a source measure unit connected to one of the aluminium fingers, keeping the cryostat temperature fixed at $20$~mK. In particular, we consider direct bias currents up to $0.35$~µA. 

We use a vector network analyzer to observe the microwave signal transmitted through the feed line of the CPW shown in Fig.\ref{fig:sample_design}a. The in-phase and quadrature components of the transmission coefficient draw a circle in the corresponding complex plane in response to the changing probe frequency of the resonator. Using a standard fitting procedure \cite{probst2015efficient}, we obtain the resonance frequency $f$ and the internal quality factor $Q_\mathrm{i}$, and thus a complete characterization of the essential state of the superconductor. 

The aluminium strip is grounded, implying that any possible gate bias current that flows through the gate electrode to the aluminum strip does not flow via the majority of the niobium resonator. Moreover, in the explored temperature range, the number of thermal quasiparticles in the niobium resonator is negligible in comparison to that in the aluminium strip due to the different critical temperatures and gap parameters: the critical temperature of niobium being $9.7$~K, the number of thermal quasiparticles is negligible up to $2$~K, well above the $1.2$~K critical temperature of aluminium. Hence, the observed change in the response of the resonator can be essentially fully attributed to the aluminium strip. Note that in this scheme, we cannot measure the fraction of the bias current possibly flowing through the aluminum strip in comparison to that leaking through the substrate. We leave such a detailed study for future work.

\subsection*{Microwave response of the resonator with temperature and gate voltage}
\begin{figure}
\centering
\begin{tikzpicture}
 \node[anchor=south west,inner sep=0] at (0,0.17\textwidth) {\includegraphics[width=0.5\textwidth]{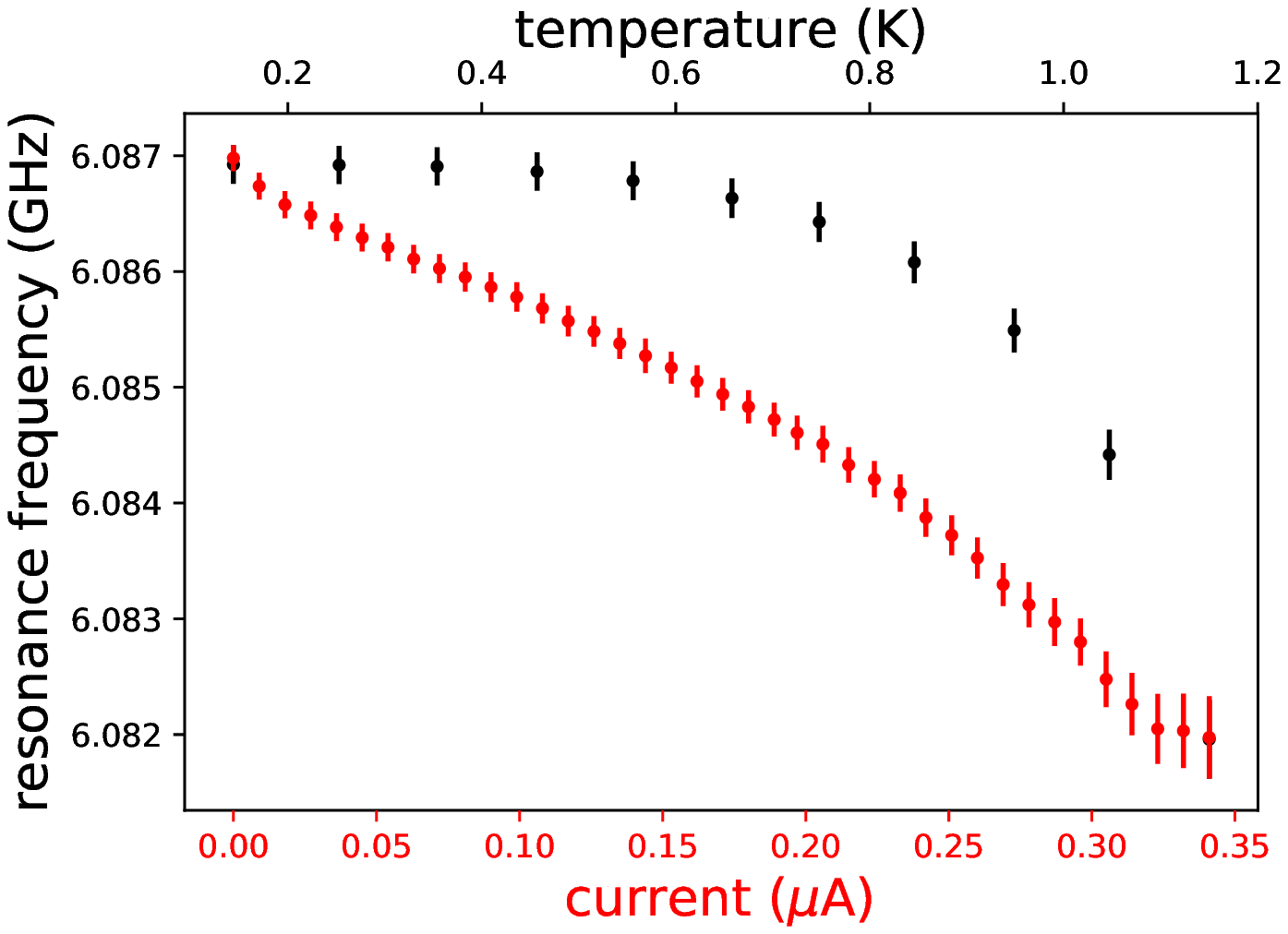}};
 \node[anchor=south west,inner sep=0] at (0,-0.17\textwidth) {\includegraphics[width=0.5\textwidth]{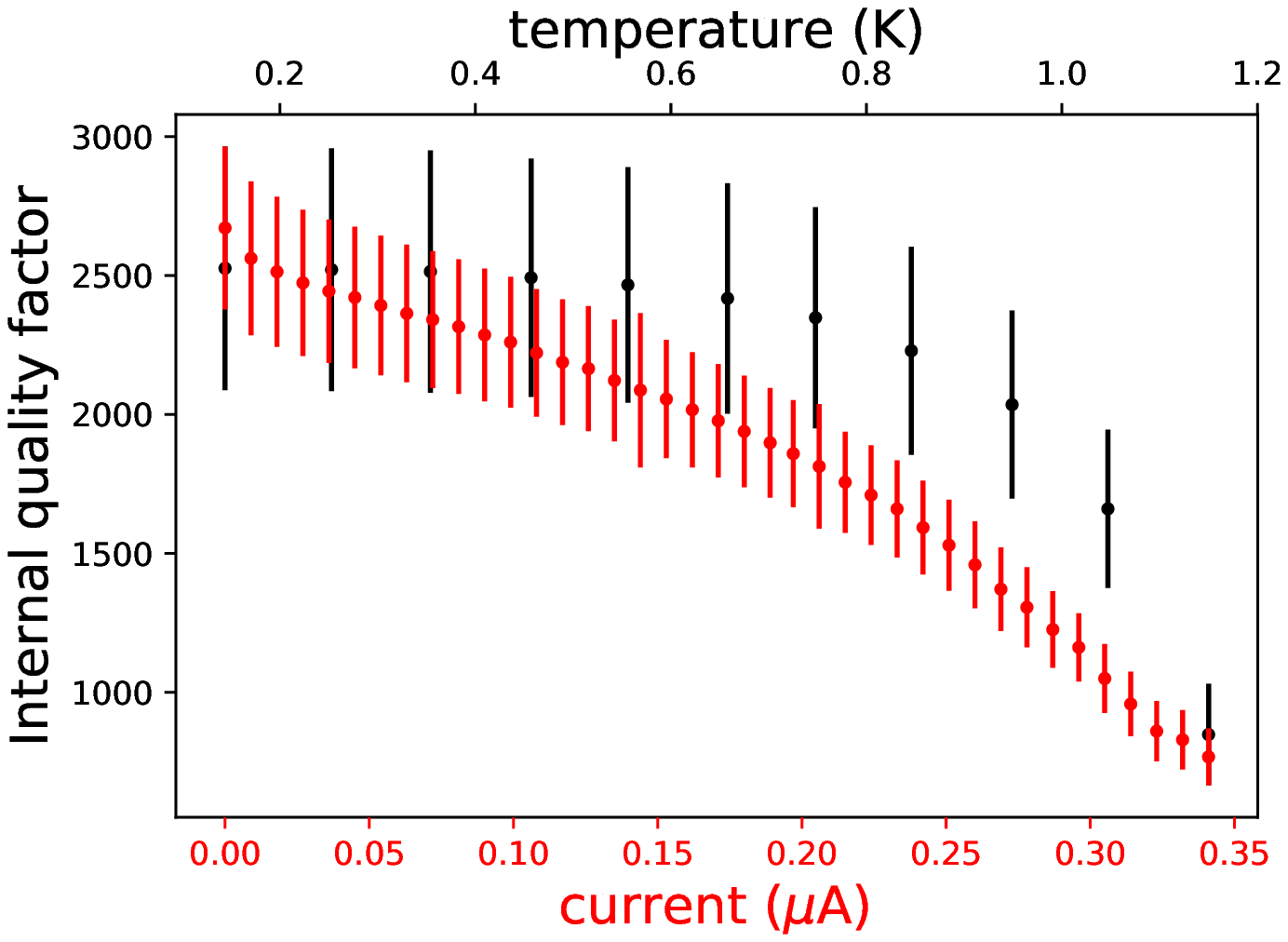}};
 \node[anchor=south west,inner sep=0] at (0.46\textwidth,0) {\includegraphics[width=0.5\textwidth]{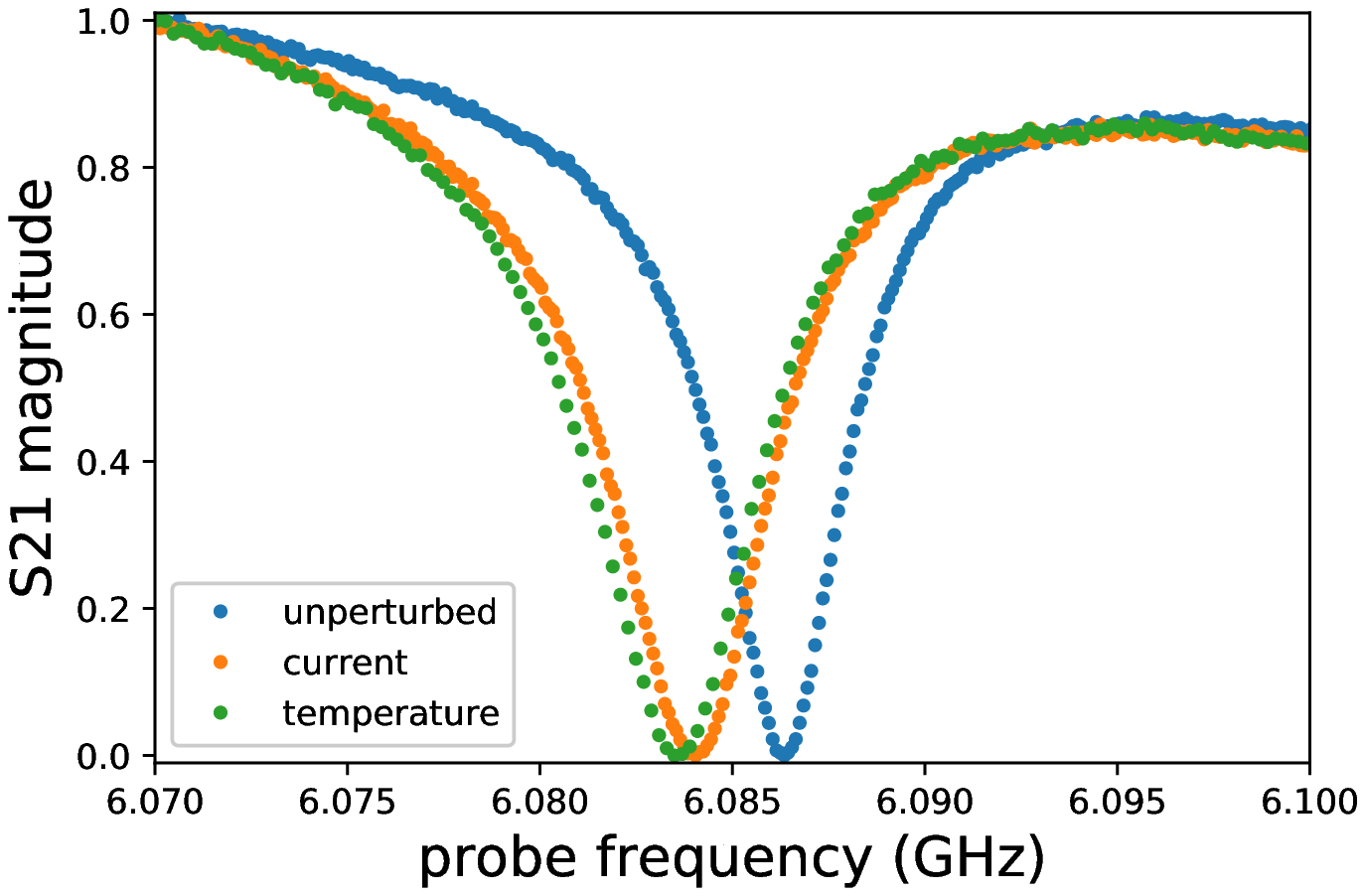}};

\node at (0.05\textwidth,0.49\textwidth) [thick, font=\fontsize{20}{20}\selectfont, thick] {(a)};
\node at (0.05\textwidth,0.15\textwidth) [thick, font=\fontsize{20}{20}\selectfont, thick] {(b)};
\node at (0.52\textwidth,0.32\textwidth) [thick, font=\fontsize{20}{20}\selectfont, thick] {(c)};
\end{tikzpicture}
\caption{(\textbf{a}) Measured resonance frequency and (\textbf{b}) internal quality factor of the resonator as functions of the bias current (red color) while keeping the bath temperature at $20$~mK and as functions of the bath temperature (black color) with no bias current. (\textbf{c}) Magnitude of the normalized transmission coefficient as a function of probe frequency under finite bias current (orange), elevated temperature (green), and not perturbations (blue).}
\label{fig:Resonance_vs_current}
\end{figure}

Figure \ref{fig:Resonance_vs_current}c shows three characteristic traces of the measured microwave transmission amplitude as a function of the probe frequency: one with no perturbation, one with elevated temperature at vanishing current, and one with finite current at 20~mK. We have chosen the value of the finite current, 180~nA, such that the corresponding trace is located close to that for the elevated temperature of 1.15~K. We observe that the perturbations tend to decrease the position of the dip in frequency and broaden the feature, in agreement with decreased resonance frequency and quality factor of the resonator. Importantly, no other significant features are visible: the shapes of the two differently perturbed traces seem essentially identical. These observations confirm our above assumption that the resonance frequency and the quality factor of the resonator are sufficient to describe the effect of the perturbations on the superconductor here. 

\begin{figure}
\centering
\begin{tikzpicture}
 
\node[anchor=south west,inner sep=0] at (0,0) {{\includegraphics[width=0.45\textwidth]{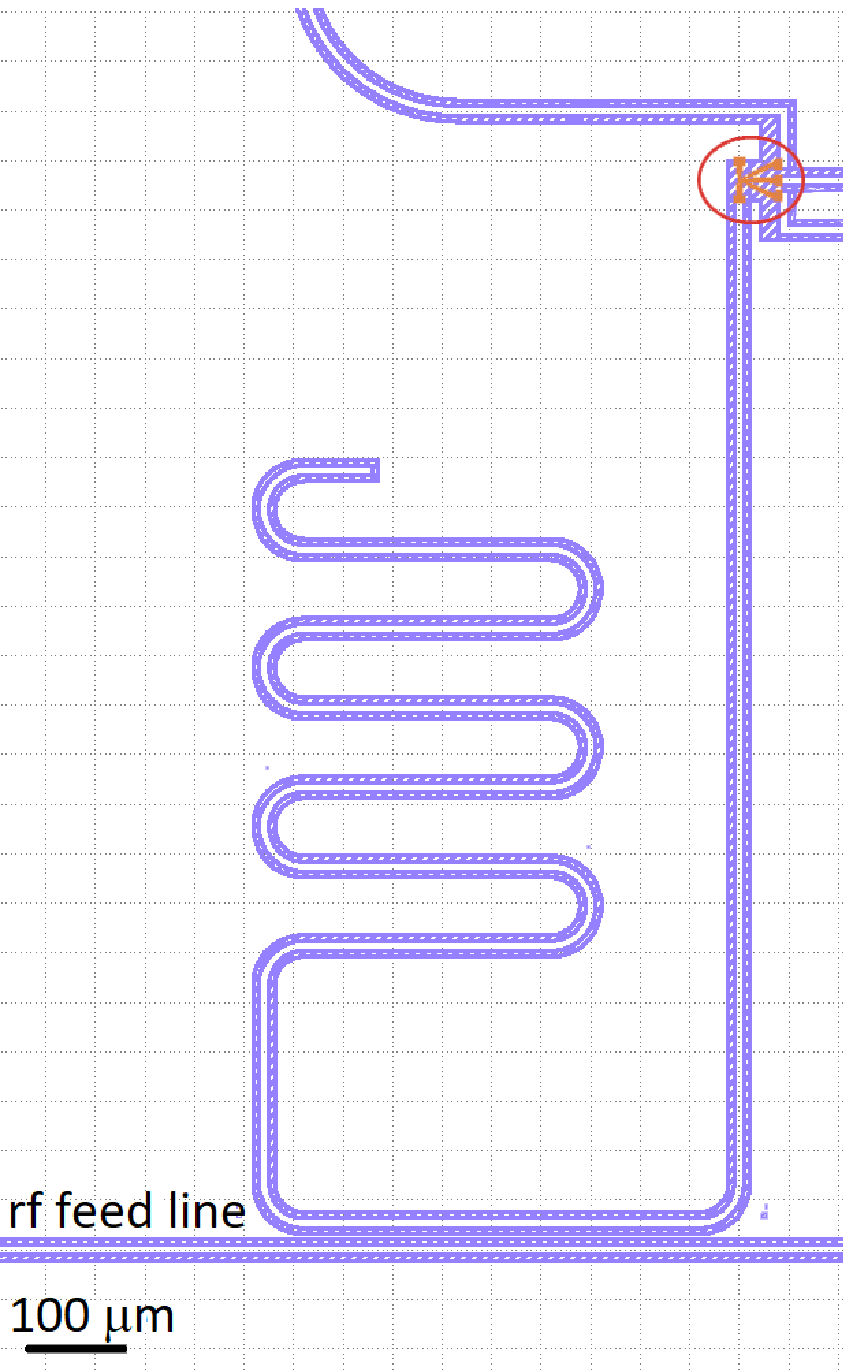}}};
\draw[black, thick] (0\textwidth,0\textwidth) rectangle (0.45\textwidth,0.725\textwidth);
\draw[red](0.42\textwidth,0.65\textwidth)--(0.495\textwidth,0.575\textwidth);
\draw[red](0.372\textwidth,0.635\textwidth)--(0.495\textwidth,0.19\textwidth);
 \node[anchor=south west,inner sep=0] at (0.495\textwidth,0.19\textwidth) {{\includegraphics[width=0.467\textwidth]{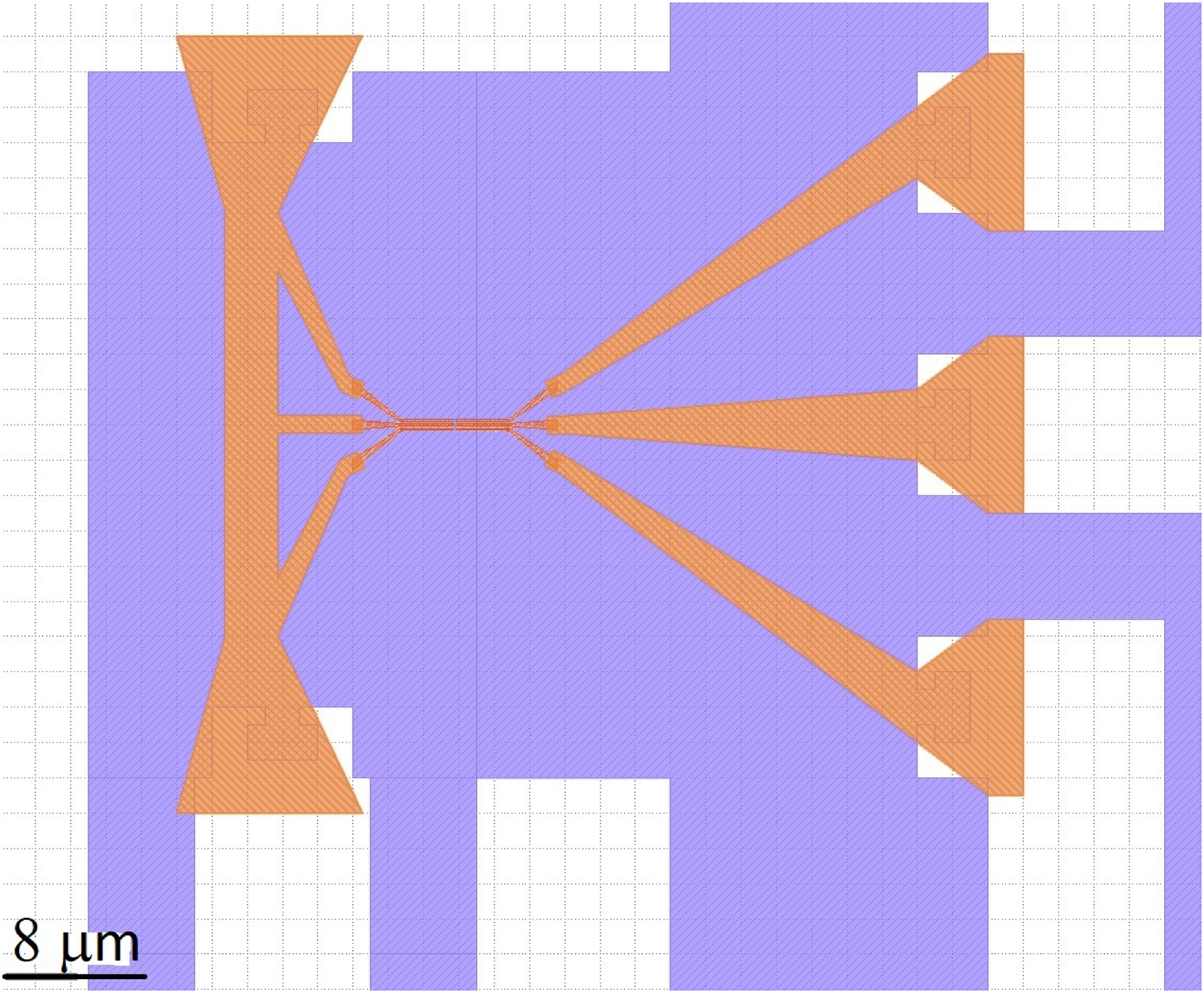}}};
 \draw[black, thick] (0.495\textwidth,0.19\textwidth) rectangle (0.96\textwidth,0.575\textwidth);
 
 \node[anchor=south west,inner sep=0] at (0.495\textwidth,-0.116\textwidth)
 {{\includegraphics[width=0.467\textwidth]{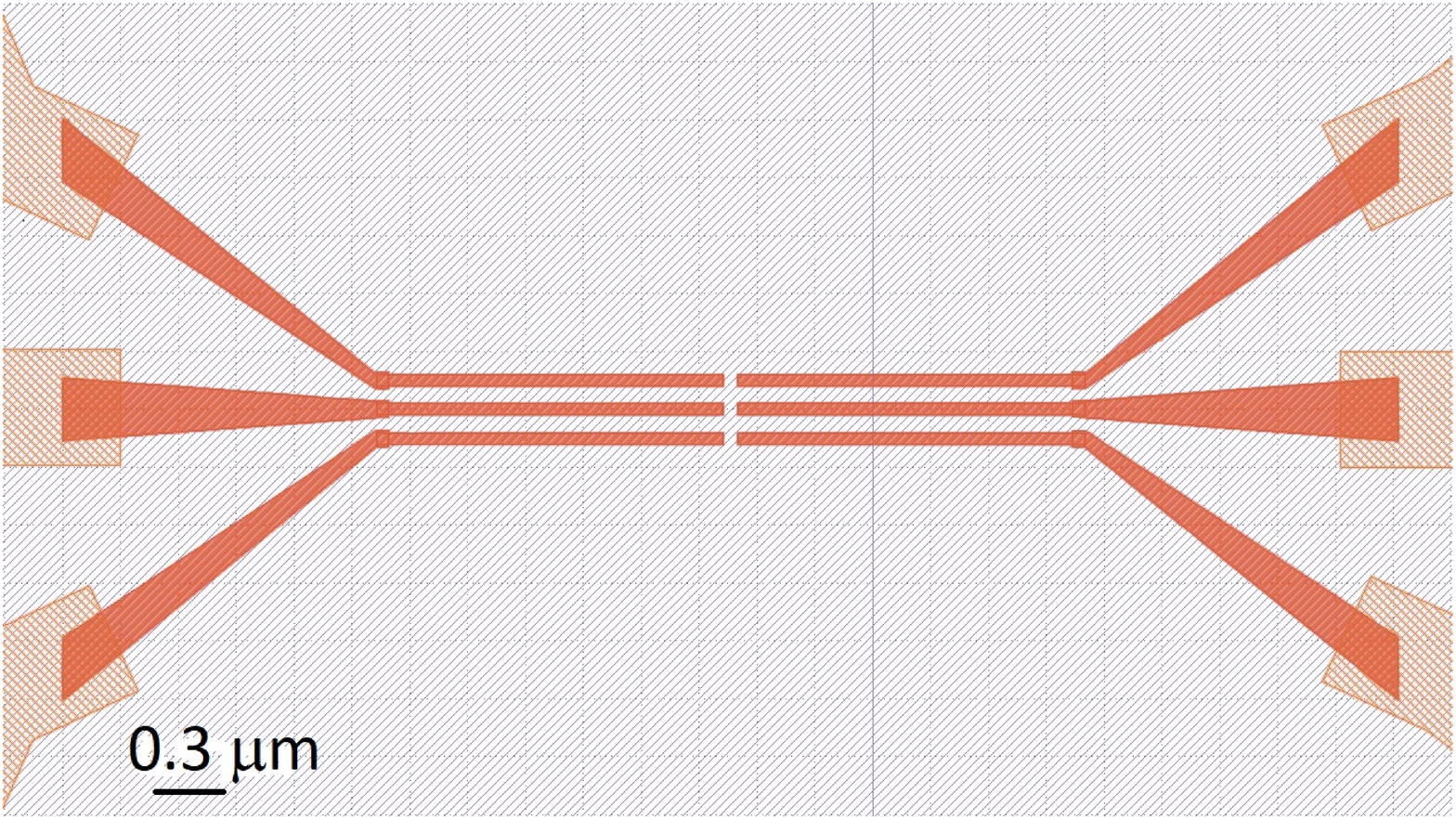}}};
 \draw[black, thick] (0.495\textwidth,-0.116\textwidth) rectangle (0.96\textwidth,0.15\textwidth);
 
 \draw[black, very thick] (0.12\textwidth,-0.116\textwidth) rectangle (0.34\textwidth,-0.017\textwidth);
 \draw[black, very thick](0.12\textwidth,-0.06\textwidth)--(-0.01\textwidth,-0.06\textwidth)--(-0.01\textwidth,0.065\textwidth)--(0.02\textwidth,0.065\textwidth);
 \draw[black, very thick] (0.34\textwidth,-0.06\textwidth)--(0.48\textwidth,-0.06\textwidth)--(0.48\textwidth,0.065\textwidth)--(0.43\textwidth,0.065\textwidth);
 \node at (0.23\textwidth,-0.07\textwidth) [thick, font=\fontsize{20}{20}\selectfont, thick] {VNA};
 
 \draw[black, very thick] (0.7\textwidth,0.59\textwidth) rectangle (0.95\textwidth,0.7\textwidth);
 \filldraw[black] (0.63\textwidth,0.643\textwidth) circle (2pt);
 \draw[black, very thick] (0.44\textwidth,0.608\textwidth)--(0.57\textwidth,0.608\textwidth);
 \draw[black, very thick] (0.44\textwidth,0.635\textwidth)--(0.57\textwidth,0.635\textwidth);

 \draw[black, very thick] (0.57\textwidth,0.67\textwidth)--(0.51\textwidth,0.67\textwidth)--(0.51\textwidth,0.735\textwidth)--(0.168\textwidth,0.735\textwidth)--(0.168\textwidth,0.718\textwidth);
 \draw[black, very thick] (0.63\textwidth,0.643\textwidth)--(0.7\textwidth,0.643\textwidth);
 \draw[black, very thick] (0.63\textwidth,0.643\textwidth)--(0.57\textwidth,0.66\textwidth);
 \node at (0.825\textwidth,0.643\textwidth) [thick, font=\fontsize{20}{20}\selectfont, thick] {SMU};
 
\draw[red] (0.675\textwidth,0.41\textwidth) ellipse (9mm and 5mm);
\draw[red](0.625\textwidth,0.41\textwidth)--(0.495\textwidth,0.15\textwidth);
\draw[red](0.728\textwidth,0.41\textwidth)--(0.96\textwidth,0.15\textwidth);

\node at (0.03\textwidth,0.7\textwidth) [thick, font=\fontsize{20}{20}\selectfont, thick] {(a)};
\node at (0.525\textwidth,0.545\textwidth) [thick, font=\fontsize{20}{20}\selectfont, thick] {(b)};
\node at (0.525\textwidth,0.13\textwidth) [thick, font=\fontsize{20}{20}\selectfont, thick] {(c)};
\end{tikzpicture}

\caption{Sample design and experimental setup. (\textbf{a}) Design of a niobium (white color) coplanar-waveguide resonator (meandering structure), capacitively coupled to an rf feed line at the bottom, connected to a vector network analyzer (VNA). The silicon substrate is indicated in shaded blue. The red ellipse highlights a small aluminium structure (orange color), used to apply the current into the resonator and to connect the resonator to the ground plane. The external current bias is implemented by a source measure unit (SMU). The SMU ground and the potential of the on-chip niobium ground plane are connected to the microwave ground of the  VNA. (\textbf{b}) Region indicated by the red ellipse in (a). The three fingers on the right constitute the gate electrodes. They are connected to the SMU via a switch using dc lines. 
(\textbf{c}) Magnified view of the current-injection structure: there is a gap of $80$ nm between the gate electrodes and the opposite fingers that are galvanically connected to the resonator.}
\label{fig:sample_design}
\end{figure}

In figs. \ref{fig:Resonance_vs_current}a and b, we show the internal quality factor of the resonator and the resonance frequency under a direct-current (dc) bias in the range from zero to 350~nA through one of the aluminium dc electrodes shown in fig.\ref{fig:sample_design}b and \ref{fig:sample_design}c. No relevant difference in the results is found between the use of one or more electrodes simultaneously with identical other parameters, suggesting that the change in the characteristics of the superconductor depends on the applied current and not independently on the applied voltage. Subsequently, we use only a single finger to apply the direct current to the aluminium strip leaving the other two fingers floating. 

After considering the perturbation by the direct current, we measure the resonance frequency and the internal quality factor of the resonator at different temperatures of the cryostat in the range from 20~mK to 1.2~K. The results shown in figs. \ref{fig:Resonance_vs_current}a and b indicate that similarly to the increasing current, the increasing temperature monotonically decreases the resonator quality factor and resonance frequency. We also observe that at the highest temperature and current point, both the resonance frequency and the internal quality factor coincide. 

\begin{figure}
\centering
\begin{tikzpicture}

 \node[anchor=south west,inner sep=0] at (0,0.17\textwidth) {\includegraphics[width=0.5\textwidth]{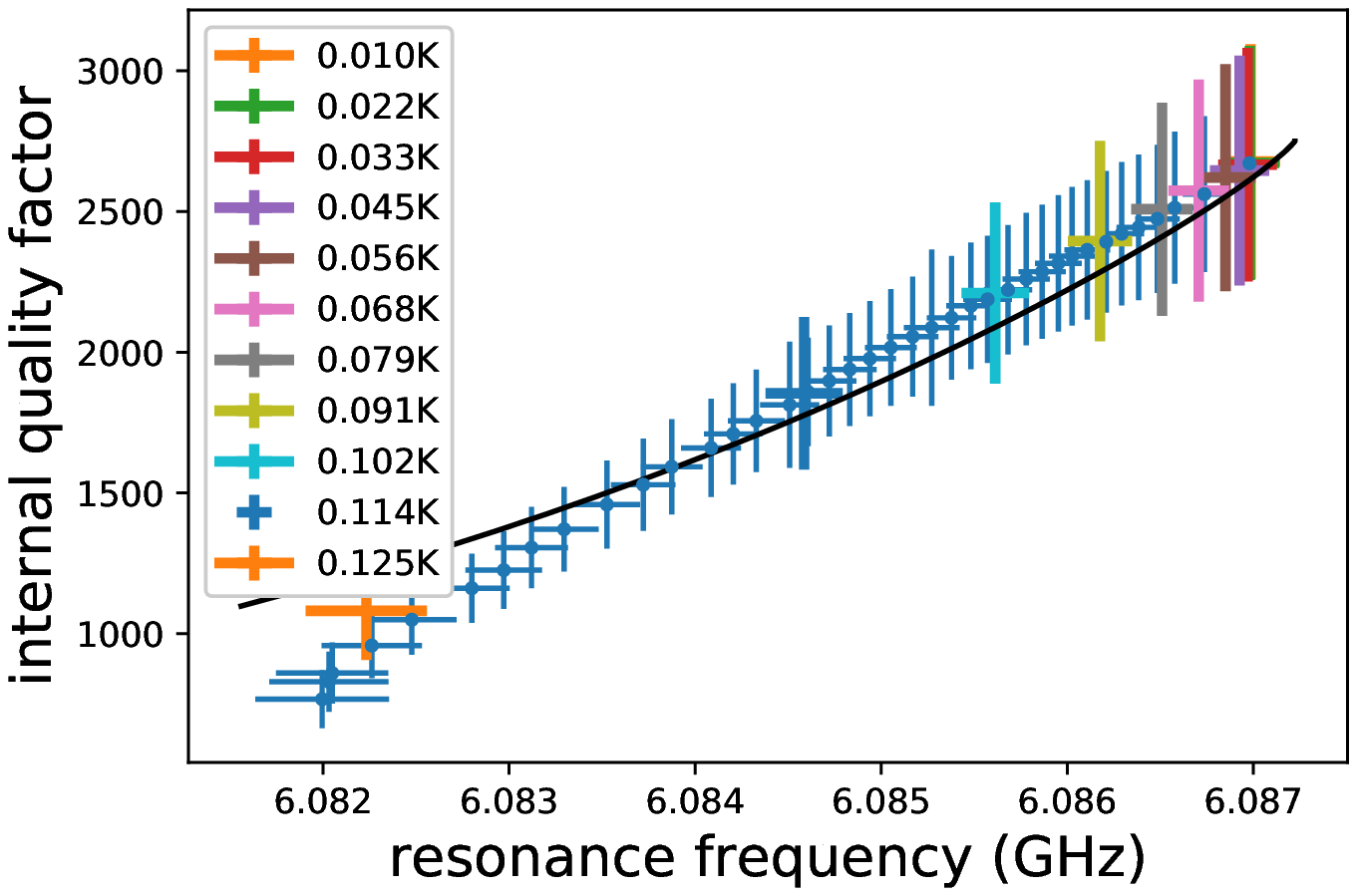}};
 \node[anchor=south west,inner sep=0] at (0,-0.17\textwidth) {\includegraphics[width=0.5\textwidth]{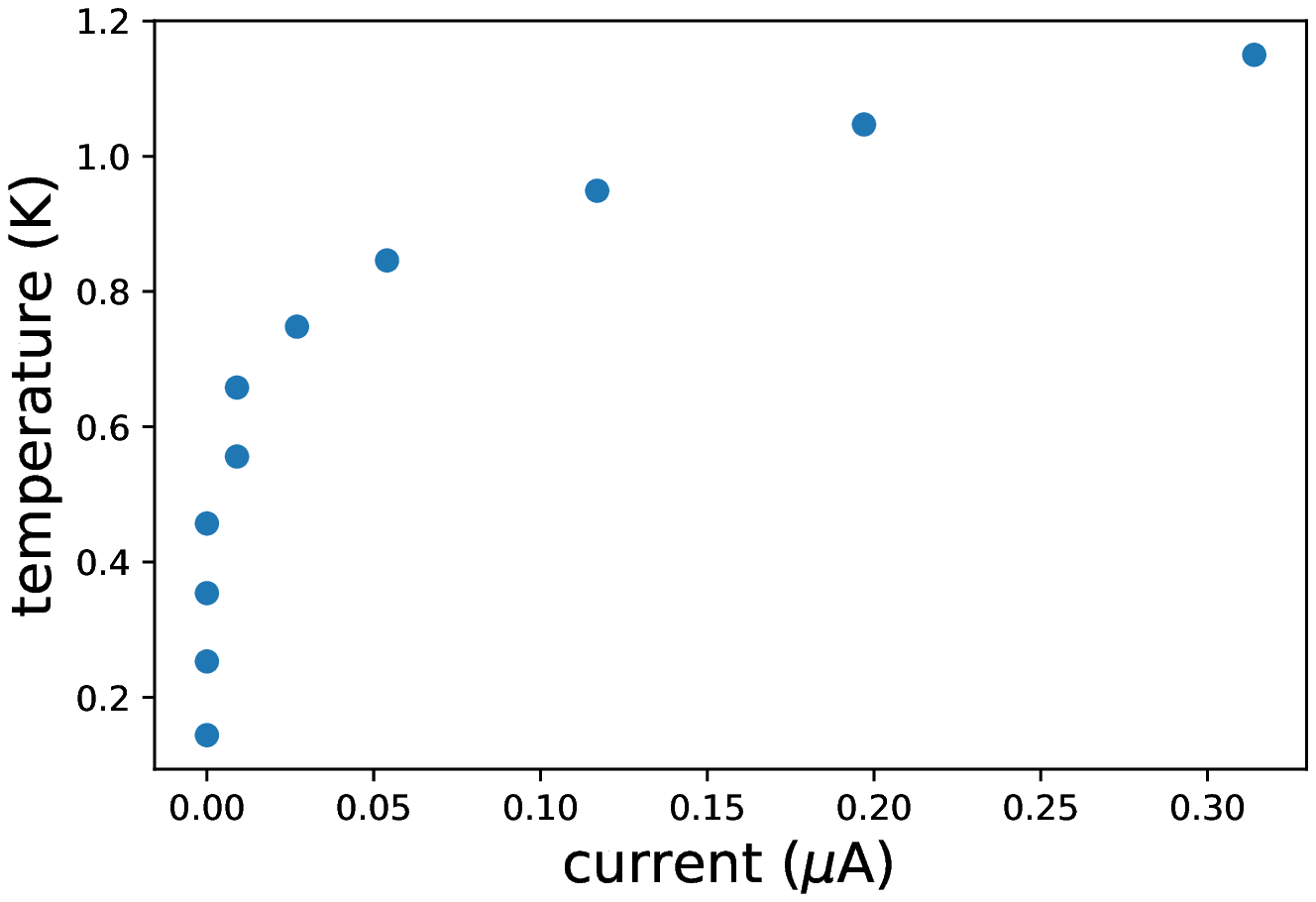}};
 \node[anchor=south west,inner sep=0] at (0.46\textwidth,0.17\textwidth) {\includegraphics[width=0.5\textwidth]{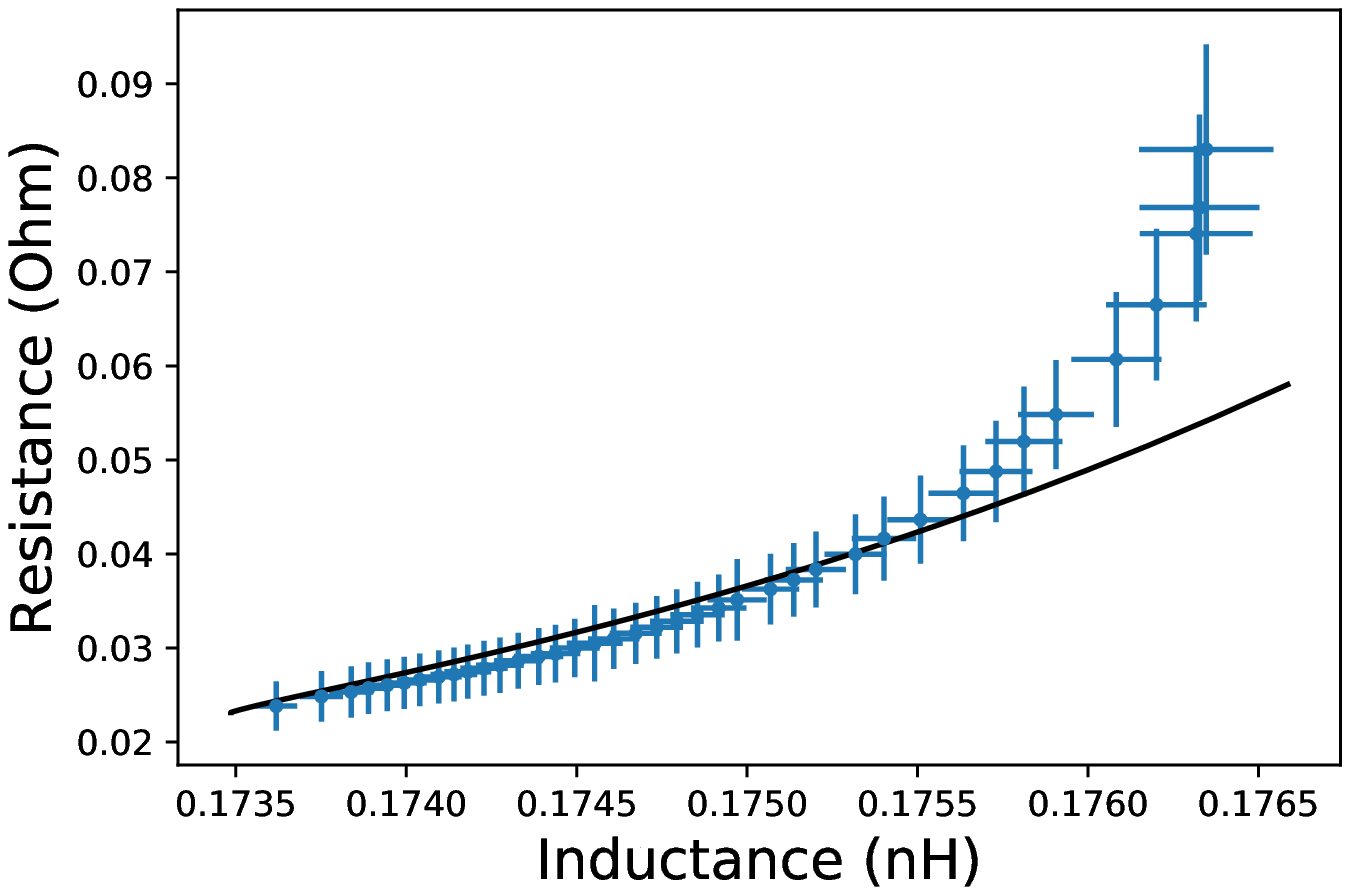}};
 \node[anchor=south west,inner sep=0] at (0.46\textwidth,-0.17\textwidth) {\includegraphics[width=0.5\textwidth]{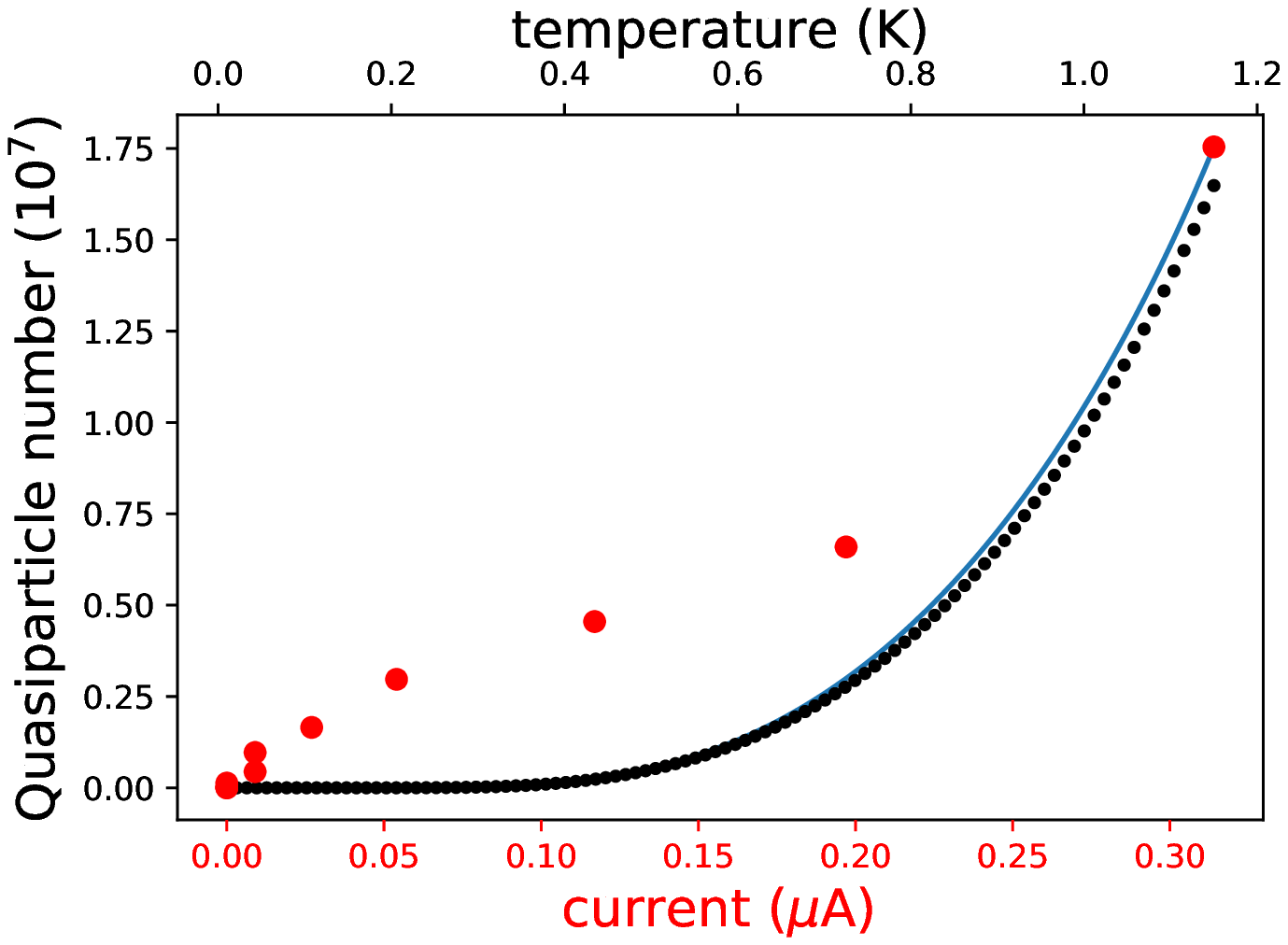}};

\node at (0.05\textwidth,0.49\textwidth) [thick, font=\fontsize{20}{20}\selectfont, thick] {(a)};
\node at (0.05\textwidth,0.16\textwidth) [thick, font=\fontsize{20}{20}\selectfont, thick] {(b)};
\node at (0.51\textwidth,0.49\textwidth) [thick, font=\fontsize{20}{20}\selectfont, thick] {(c)};
\node at (0.51\textwidth,0.16\textwidth) [thick, font=\fontsize{20}{20}\selectfont, thick] {(d)};

\end{tikzpicture}
\caption{\textbf{(a)} 
Internal quality factor and resonance frequency of the resonator at different values of the applied direct current (blue markers), from $0$~A (highest values) to $0.35$~µA (lowest values), and at different cryostat temperatures as indicated by the colors. The black solid line is obtained from the Mattis--Bardeen theory as described in the Methods. (\textbf{b}) Temperature of the resonator as a function of the applied current. We find these data points by matching the the data for different bias currents in (a) with those recorded at different temperatures. \textbf{(c)} Effective inductance and resistance of the aluminium strip (markers) according to the lumped-element model at different perturbation currents, from the unperturbed state (bottom left) to the $0.35$~µA perturbation (top right). The data has been extracted from (a) considering a constant effective impedance contribution of 1.5 nH and a negligible effective resistance for the bare resonator. The solid line provides the corresponding values from the results of the Mattis--Bardeen theory in (a). \textbf{(d)} Number of quasiparticles in the aluminum strip  as a function of the temperature (top horizontal axis) and bias current (bottom horizontal axis). The blue solid line considers the temperature dependence (top horizontal axis) from the integral in equation~\eqref{n_q} and the black dots correspond to the approximation provided at the end of equation~\eqref{n_q}. The red points show the number of quasiparticles as a function of the bias current (bottom horizontal axis) according to the integral form of equation~\eqref{n_q} where the temperature at a given bias current is obtained from panel (b).}
\label{fig:Q_vs_Resonance}
\end{figure}

In fig.\ref{fig:Q_vs_Resonance}a, we study the measured internal quality factor as a function of the measured resonance frequency under the chosen perturbations. This study reveals that within the experimental uncertainty, the response of the resonator to the bias current is identical to that at different temperatures. Thus, we can neglect any possible field-effect response of the resonator and describe the state of the superconductor under an external current by simply considering a change of its temperature, behavior that is also in relatively good agreement with the Mattis--Bardeen theory (see Methods) as illustrated in the figure. Utilizing this identification, we extract the effective quasiparticle temperature of the resonator as a function of the applied current as shown in fig. \ref{fig:Q_vs_Resonance}b. As expected from the low-temperature decay of the thermal quasiparticle density following equation~\eqref{n_q}, the temperature sensitivity of our signal quickly decreases with temperature decreasing well below the critical temperature of aluminum (see also figs.~\ref{fig:Resonance_vs_current}a and \ref{fig:Resonance_vs_current}b), and consequently we restrict our temperature scale from below to 144~mK. Notably, the extracted temperature increases rapidly even for a current of 10~nA. Following the lumped-element model of equations~\eqref{resonanceFrequency} and ~\eqref{internalQ}, in fig.\ref{fig:Q_vs_Resonance}c we show the resistance and the inductance of the system at different values of the perturbation current. 

In fig. \ref{fig:Q_vs_Resonance}d, we present the expected number of quasiparticles in the aluminium strip according to equation~ \eqref{n_q} and observe that the numerical integral provides more accurate values near the critical temperature than the low-temperature analytical approximation. In addition, we convert the used bias currents into a temperature scale with the help of fig.~\ref{fig:Q_vs_Resonance}b and subsequently use the integral form of equation~ \eqref{n_q} to extract the expected number of quasiparticles as a function of the the applied current.

\section*{Conclusions}

We designed and fabricated a superconducting coplanar-waveguide resonator which is grounded from one side by an aluminum strip. By extracting the resonance frequency and internal quality factor of the resonator, we tracked the state of the superconducting aluminium strip at different cryostat temperatures and different bias currents from nearby gate electrodes. From a detailed comparison of these two data sets, namely, from the fact that each resonator quality factor at an elevated temperature equals that at a gate leakage current tuned to yield an equal resonance frequency, we conclude that the change in the reactance and resistance of the superconductor in our device can be well described simply by heating of the aluminum strip. Any possible field effect that the external voltage may cause seems, in our case, negligible in comparison to the resulting Joule heating created by the injected current. In addition, our observations are in a relatively good agreement with Mattis--Bardeen theory of a heated superconductor. The change of the quasiparticle temperature of the superconductor, induced by the gate leakage current driven by the electric field, well describes the phenomenon. Note that a possible field effect itself is not expected to excite quasiparticles but to decrease the gap of the superconductor and hence the resonance frequency of the resonator with no major effect on the internal quality factor. 

Consequently, our result agrees with the recently published work\cite{golokolenov2021origin} on the subject, where the field-effect hypothesis was not required, which points to the need of additional care in the design and fabrication of devices that are desired to pronounce such an effect. Whereas the recent work\cite{golokolenov2021origin} studied a case where the gate electrodes induced an electric field in the vicinity of the point through which the resonator was grounded, we in contrast have the electrodes relatively far away from the grounding point. On one hand, this different geometry is expected to greatly decrease the visibility of any possible field effect in our experiment, but on the other hand, it shows that the heating effect can be also captured from distance. Note we do not claim our experiments to nullify the results of the previous works owing to these differences and also owing to the fact that we cannot measure in this setup the fraction of the bias current flowing through the aluminum strip in comparison to that leaking through the substrate. Interestingly, our gate configuration allows for some level of control on the distribution of the resulting electric field provided that one independently biases each gate. This feature, together with the choice of insulating substrates, may be used in the future to optimize the fraction of ejected quasiparticles captured by the strip.

In addition to the connection of our work to the fundamental knowledge, the considered scheme may be studied in the future for applications. In principle, the device can be used as both a current sensor and as a secondary thermometer: the utilized material determines the working temperature range, whereas the design determines the precision in the measurement of current. We leave the estimation of these properties for future work.

\section*{Methods}

\subsection*{Sample fabrication}
The device were fabricated at Micronova cleanroom.  First, a high-resistivity ($\rho$  > 10 k$\Omega$ cm) non-oxidised $n$-type-undoped (100) 4-inch silicon wafer is sputtered by 200 nm of highly pure Nb. Then, AZ5214E photo resist in positive mode with hard contact is applied in photolithography to form the coplanar waveguide (CPW) using a Karl Suss MA-6 mask aligner. After development, the Nb film is etched with Plasmalab 80Plus Oxford Instruments reactive ion etching (RIE) system. The plasma works in a gas flow of $\mathrm{SF_{6}}$/$\mathrm{O_{2}}$ at 40~sccm/20~sccm with a rf-field power of 100~W. After etching, the resist residuals are cleaned in ultrasonic bath with acetone, IPA, and dried with a nitrogen gun. Subsequently, the 4-inch wafer is cleaved into a 3 $\times$ 3 cm$^{2}$ chip by Disco DADdy, including 9 pixel chips in total. Each pixel chip is 1 $\times$ 1 cm$^{2}$. 

The superconducting electrodes are patterned by a 100-keV EPBG5000pES electron beam lithography (EBL) system with a bilayer of PMMA 495 A4/PMMA 950 A4 resist on a single chip. The optimal dose is 1100 $\mu$C/cm$^{2}$. Proximity effect correction (PEC) in Beamer is used to ensure the split gate gap to vary only within a few nanometers. This is followed by a development in MIBK:IPA (1:3) 40 s and IPA 20 s. The resist residues are cleaned with oxygen descum of 15 s. Al is deposited inan electron beam evaporator at a rate of 1.5 Å/s. Before evaporation, the native oxides is removed by Ar ion milling. The fine feature of the split gate is 30-nm thick and the large electrode is 60-nm thick. Finally, each pixel is cleaved by a laser micromachining system, packaged, and bonded with Al wires.

\subsection*{Simulation}
To predict the effective values of the resistance and inductance shown in fig.~\ref{fig:Q_vs_Resonance}c, we utilized a simple lumped element model. However, in order to reliably simulate the resonance frequency, quality factor, and kinetic inductance, a more refined model is needed. To this end, we employ the Mattis--Bardeen theory and simulate the sheet impedance of the aluminum strip as a function of temperature\cite{de2014quasiparticle}. The underlying aluminum parameters are $\Delta(0) = 200~\mu\mathrm{eV}$ and a normal-state resistivity of $7.2\times 10^{-14}~\Omega~\mathrm{nm}$. The superconducting gap is assumed to follow a temperature dependence given by the Bardeen-Cooper-Schrieffer theory. The Mattis-Bardeen sheet impedance is passed to a script that numerically minimizes the sum of the kinetic energy and the magnetic field energy in a cross-sectional transmission line geometry that combines the aluminum strip with a niobium ground plane. Within the script the superconductors are infinitesimally thin, but the estimated degree of magnetic field penetration into the 90-nm-thick aluminum is however considered in its sheet impedance. We find that the geometric inductance dominates the resulting impedance, and that the inhomogeneity of the current density across the cross section makes the kinetic inductance slightly larger than the prediction of equation~\ref{kineticInductance}.

We then embed the aluminum strip impedance into the CPW resonator. Following the theory of coupled transmission lines \cite{PozarBook} for the section of the resonator parallel to the feedline, we calculate the even and odd mode characteristic impedances \cite{martinis2014calculation} and apply the $S$-parameter matrix associated to the coupler to derive how the resonator modifies the feedline microwave transmission. Analogously with the experiment, the simulated transmission shows a resonance dip, and we extract the resonance frequency and quality factor from the best fit to the complex-valued transmission S-parameter. Since the microwave loading of the resonator by the SMU circuitry is not captured by the model, we use the length of the resonator as a free parameter and introduce a parallel, temperature-independent loss channel to fix the low-temperature resonance frequency and quality factor to the experimental values. As a result, the obtained curves that describe the state of the resonator fall within the uncertainty of our measurements at temperatures below 1.1~K as shown in fig.~\ref{fig:Q_vs_Resonance}a.

\bibliography{sample}

\section*{Acknowledgements}
We have received funding from European  Research Council under Grant No~681311 (QUESS) and Academy of Finland through its Centres of Excellence Programme (project numbers ~312300, ~312295 and 336819) and grants (numbers 314449, 321700 and 335460). We thank Yuri Pashkin for useful discussions. We acknowledge the provision of facilities and technical support by Aalto University at OtaNano -- Micronova Nanofabrication Center. 

\section*{Author contributions statement}
G.C. conducted the experiment and data analysis. The sample was designed and fabricated by W.L. S.K. managed the experimental setup and the dc heater of the cryostat. V.L. assisted in the interpretation of the results. V. V. developed the simulation methods. The work was actively supervised by M.M. The manuscript was written by G.C., V.L., W.L. and M.M. with input from all authors.

\section*{Competing interests}
The authors declare no competing interests.

\newpage

\end{document}